\documentclass[oneside,reqno,a4paper,12pt]{amsart}

\usepackage{latexsym}
\usepackage{euscript}

\newcommand{\mathscript}{\EuScript}
\numberwithin{equation}{section}

\newcommand{\dud}[4]{{\smash{{{#1}_{#2}\/}^{#3}}}_{#4}}
\newcommand{\du}[3]{{\smash{{#1}_{#2}\/}^{#3}}}
\newcommand{\ud}[3]{{\smash{{#1}^{#2}\/}_{#3}}}
\newcommand{\sud}[3]{\smash{#1^{#2}_{#3}}}

\newcommand{\SU}{\mathrm{SU}}

\newcommand{\cc}{s}
\newcommand{\tr}{\,\mathrm{tr}\,}

\makeatletter
    \def\serieslogo@{\vtop to 
0pt{\noindent\scriptsize\ppn\parindent\z@}}
    \let\@setcopyright\@empty
\makeatother

\begin{document}

\def\ppn{HEP-TH/9706216, UPR-764T}

\title[Two-Form Formulation of the Vector-Tensor Multiplet]{A Two-Form 
Formulation of the Vector-Tensor\\
Multiplet in Central Charge Superspace}

\maketitle

\begin{center}
\textsc{Ioseph Buchbinder} \\
\vspace*{1mm}
\small{\textit{Department of Theoretical Physics,
Tomsk State Pedagogical University,\\
Tomsk 634041, Russia}} \\
\vspace*{5mm}
\textsc{Ahmed Hindawi} \\
\vspace*{1mm}
\small{\textit{Department of Physics, Faculty of Science, Ain Shams 
University,\\
Cairo 11566, Egypt}} \\
\vspace*{5mm}
\textsc{Burt A. Ovrut} \\
\vspace*{1mm}
\small{\textit{Department of Physics, University of Pennsylvania,\\
Philadelphia, PA 19104--6396, USA}\\
\vspace*{1mm}
and\\
\vspace*{1mm}
\textit{Institut fur Physik, Humboldt-Universitat,\\
Invalidenstrasse 110, D-10115 Berlin, Germany}} \\
\end{center}

\vspace*{5mm}

\begin{abstract}

A two-form formulation for the $N=2$ vector-tensor multiplet is constructed
using superfield methods in central charge superspace. The 
$N=2$ non-Abelian standard supergauge multiplet in central charge 
superspace is also discussed, as is with the associated Chern-Simons form. 
We give the constraints, solve the Bianchi identities and present the action
for a theory of the vector-tensor multiplet coupled to the non-Abelian 
supergauge multiplet via the Chern-Simons form.

\vspace*{\baselineskip}

\noindent PACS numbers: 11.30.Pb, 11.15.-q

 
\end{abstract}

\renewcommand{\baselinestretch}{1.2} \large\normalsize

\vspace*{\baselineskip}

\section{Introduction}

In a recent paper \cite{PLB-392-85}, we constructed the $N=2$ vector-tensor 
supermultiplet \cite{PLB-92-123,NPB-173-127} as a constrained supergauge 
multiplet using superfield techniques in $N=2$ central charge superspace 
\cite{PLB-392-85,PLB-373-89,Gaida96b}. In that work, the gauge component 
field of this supermultiplet was introduced via an Abelian gauge connection 
superfield and the antisymmetric tensor component field derived as a 
consequence of a particular choice of constraints. In \cite{PLB-392-85}, we 
presented the theory for a free, non-interacting vector-tensor multiplet, 
preferring to concentrate on the important theoretical issues, such as the 
choice of constraints, the central charge hierarchy and construction of a 
supersymmetric action. Our work, however, was motivated by recent results in
superstring theory \cite{NPB-451-53}, where the interacting vector-tensor 
multiplet emerged as being of fundamental importance. The coupling of the 
vector-tensor multiplet to supergravity and to vector multiplets has also 
recently been considered \cite{PLB-373-81,HEP-TH/9612203}. It was shown that 
this requires the gauging of central charge, leading to a Chern-Simons 
coupling between the vector-tensor multiplet and the $N=2$ standard vector 
multiplet. With this in mind, it is the purpose of this paper to present the 
general theory for the coupling of the $N=2$ vector-tensor supermultiplet to 
both Abelian and non-Abelian $N=2$ standard supergauge multiplets via the 
associated super Chern-Simons forms. To do this, it is convenient to present 
an alternative construction for the vector-tensor multiplet as a constrained 
super two-form. We do this using superfield techniques in $N=2$ central 
charge superspace. In this approach, the two-form component field of this 
supermultiplet is introduced via a two-form superfield and the gauge 
component field derived as a consequence of a particular choice of 
constraints. We will also present, in central charge superspace, the general 
theory for the $N=2$ non-Abelian standard supergauge multiplet and its 
associated Chern-Simons form. Finally, we can combine these two formalisms 
to construct the general central charge superspace theory for the coupling 
of the vector-tensor supermultiplet to an Abelian or non-Abelian standard 
supergauge multiplet via the Chern-Simons form. We explicitly give the 
constraints, solve the Bianchi identities and give the superfield action for 
this coupled theory.

\section{$N=2$ Central Charge Superspace}

As discussed in \cite{NPB-138-109,NPB-133-275,PLB-392-85}, $N=2$ central 
charge superspace is a space with coordinates $z^M = 
(x^m, \theta^\alpha_i, \bar\theta_{\dot\alpha}^i, z, \bar z)$ where $x^m$, 
$z$, and $\bar z=z^*$ are commuting bosonic coordinates while 
$\theta^\alpha_i$ and $\bar\theta_{\dot\alpha}^i=(\theta_{\alpha i})^*$ 
are anticommuting fermionic coordinates. Taylor-series 
expansion of a generic superfield in the $\theta$ coordinates 
terminates after a finite number of terms due to the anticommuting 
nature of $\theta$. On the other hand, the expansion in $z$ and $\bar 
z$ never ends. This means there are, a priori, an infinite 
number of component fields in a general superfield. These can, however, be 
reduced to a finite number off-shell by applying appropriate constraints. 

Translations in this superspace are generated by the supercovariant 
differential operators $\partial_a$, $\partial_z$, $\partial_{\bar z}$, 
and
\begin{equation}
\begin{split}
D_\alpha^i &= \frac{\partial}{\partial \theta^\alpha_i} + i \ud\sigma a 
{\alpha\dot\alpha} \bar\theta^{\dot\alpha i} \partial_a - i 
\theta_\alpha^i \partial_z, \\
\bar D_{\dot\alpha i} &= - \frac{\partial}{\partial 
\bar\theta^{\dot\alpha i}} - i \theta^\alpha_i \ud\sigma a 
{\alpha\dot\alpha} \partial_a - i \bar\theta_{\dot\alpha i} 
\partial_{\bar z},
\end{split}
\end{equation}
where $\bar D_{\dot\alpha i}=-(D^i_\alpha)^\dag$. The anticommutation 
relations for these operators are
\begin{equation}
\begin{split}
\{ D_\alpha^i , \bar D_{\dot\alpha j} \} &= - 2i \ud\delta i j \ud 
\sigma m {\alpha\dot\alpha} \partial_m, \\
\{ D_\alpha^i , D_\beta^j \} &= -2i \epsilon_{\alpha\beta} 
\epsilon^{ij} \partial_z, \\
\{ \bar D_{\dot\alpha i} , \bar D_{\dot\beta j} \} &= 2i 
\epsilon_{\dot\alpha\dot\beta} \epsilon_{ij} \partial_{\bar z}.
\end{split}
\label{Dalg}
\end{equation}

The supervielbein $\du e A M$ of this superspace is defined as the 
matrix that relates the supercovariant derivatives $D_A=(\partial_a, 
D^i_\alpha, \bar D_{\dot \alpha i}, \partial_z, \partial_{\bar z})$ and 
the ordinary partial derivatives
\begin{equation}
D_A = \du e A M \frac\partial{\partial z^M}.
\end{equation}
The torsion $T^A$ 
is defined as
\begin{equation}
T^A = d e^A  = \tfrac12 e^C e^B \du T {BC} A.
\end{equation}
The non-vanishing components of the torsion are found to be
\begin{equation}
\begin{split}
\smash{\smash{T^i_\alpha}^j_{\dot\alpha}}^a &= {\sud{\sud T j 
{{\dot\alpha}}} i \alpha}^a = - 2 i \epsilon^{ij} \ud\sigma a 
{\alpha\dot\alpha}, \\
\sud{\sud T i \alpha}j\beta^z  &= \sud{\sud T j\beta} i \alpha^z = 2 i  
\epsilon^{ij} \epsilon_{\alpha\beta}, \\
\sud{\sud T i {\dot\alpha}} j {\dot\beta}^{\bar z} &= \sud{\sud T j 
{\dot\beta}} i {\dot\alpha}^{\bar z} =  2 i \epsilon^{ij} 
\epsilon_{\dot\alpha\dot\beta}. 
\end{split}
\end{equation}

\section{Vector-Tensor Multiplet: Gauge Theory Formulation}

In this section we briefly describe the super-gauge theory formulation of 
the vector-tensor multiplet presented in \cite{PLB-392-85}. We begin by 
considering the geometrical form of super-gauge theory in superspace with 
central charge. We then find a suitable set of constraints on the super 
field strengths to reproduce the field content of the vector-tensor 
multiplet. Here we restrict our attention to Abelian gauge theories, 
returning to a discussion of non-Abelian theories below. Let us introduce an 
hermitian connection $A=dz^M A_M = e^A A_A$. The hermiticity of the 
connection implies
\begin{equation}
A = dx^a A_a + d\theta^\alpha_i A^i_\alpha + d\bar\theta^i_{\dot\alpha} 
\bar A_i^{\dot\alpha} + dz A_z + d\bar z A_{\bar z},
\end{equation}
where $A_a$ is real, $\bar A^{\dot\alpha}_i = (A^{\alpha i})^\dag$, and 
$A_{\bar z} = A_z^\dag$. The curvature two-form is defined as
\begin{equation}
F = dA = \tfrac12 e^B e^A F_{AB}.
\end{equation}
The curvature tensor $F$ is subject to the Bianchi identities $dF=0$.
It is natural to adopt a set of constraints 
which set the pure spinorial part of the curvature tensor to zero. That 
is 
\begin{equation}
\sud{\sud F i \alpha} j \beta = \sud{\sud F i \alpha} j {\dot\beta} = 
\sud{\sud F i {\dot\alpha}} j {\dot\beta} = 0.
\label{con-1}
\end{equation}
We would like to explore the 
consequences of \eqref{con-1}. To do so we must solve the Bianchi 
identities subject to these constraints. The result is that all the 
components of the curvature tensor $F_{AB}$ are determined in terms of 
a single superfield $F^i_{\alpha\bar z}$ and its hermitian conjugate 
$F_{\dot\alpha i z} = (F^i_{\alpha \bar z})^\dag$. Henceforth, we 
denote these superfields by $W^i_\alpha$ and $\bar W_{\dot\alpha i}$. 
In particular $F^i_{\alpha z} = 0$ and
\begin{align} 
F_{ab} &= -\tfrac1{16} i \epsilon_{ij} 
\du{\bar\sigma}a{\dot\alpha\alpha} \du{\bar\sigma}b{\dot\beta\beta} 
\left( \epsilon_{\dot\alpha\dot\beta} D^j_\beta W^i_\alpha + 
\epsilon_{\alpha\beta} \bar D^j_{\dot\beta} \bar W^i_{\dot\alpha} 
\right). \label{3.32} \\
F^i_{\dot\alpha a} &= - \tfrac12 \epsilon_{\dot\alpha\dot\beta} 
\du{\bar\sigma}a{\dot\beta\alpha} W^i_\alpha, \label{3.31} \\ 
F_{a\bar z} &= - \tfrac18 i \epsilon_{ij} 
\du{\bar\sigma}a{\dot\beta\alpha} \bar D^j_{\dot\beta} W^i_\alpha, 
\label{3.30} \\
F_{z\bar z}  &= \tfrac14 i \epsilon_{ij} \epsilon^{\alpha\beta} 
D_\alpha^i W_\beta^j, \label{3.29}
\end{align}
Furthermore, $W^i_\alpha$ is constrained to satisfy 
\begin{align}
& D^{(j}_{(\beta} W^{i)}_{\alpha)} = 0, \qquad \bar D^{(j}_{(\dot\beta} 
\bar W^{i)}_{\dot\alpha)} = 0, \label{1} \\
& \bar D^{(j}_{\dot\beta} W^{i)}_{\alpha} = 0, \qquad D^{(j}_\beta \bar 
W^{i)}_{\dot\alpha} = 0, \label{2} \\
& \epsilon^{\alpha\beta} D_\alpha^{[i} W_\beta^{j]} 
= - \epsilon^{\dot\alpha\dot\beta} \bar D_{\dot\alpha}^{[i} \bar 
W_{\dot\beta}^{j]}, \label{3} \\
& \epsilon^{\alpha\beta} D_\alpha^{(i} W_\beta^{j)} 
= \epsilon^{\dot\alpha\dot\beta} \bar D_{\dot\alpha}^{(i} \bar 
W_{\dot\beta}^{j)}.
\label{4}
\end{align}

Let us explore the consequences of \eqref{1}--\eqref{4} on the field 
content of $W^i_\alpha$. The expansion of $W^i_\alpha$ in the 
anti-commuting coordinates has the general form
\begin{equation}
W^i_\alpha = \lambda^i_\alpha + \theta_j^\beta \sud{\sud G i \alpha} j 
\beta + \bar\theta_j^{\dot\alpha} \sud{\sud H i \alpha} j {\dot\alpha} 
+ \mathscript{O}(\theta^2).
\label{c1}
\end{equation}
Conditions \eqref{1}--\eqref{4} will not impose any 
restriction on $\lambda^i_\alpha$. First we consider the implications 
of the lowest component of superfield constraints \eqref{1}--\eqref{4}. 
Condition \eqref{1} implies that
\begin{equation}
\sud{\sud G i \alpha} j \beta = i \epsilon^{ij} f_{\alpha\beta} 
+ 2\epsilon^{ij}\epsilon_{\alpha\beta} D + i \epsilon_{\alpha\beta} 
\rho^{ij},
\label{c2} 
\end{equation}
where $f_{\alpha\beta}=f_{(\alpha\beta)}$ and $\rho^{ij}=\rho^{(ij)}$. 
Conditions \eqref{3} and \eqref{4} further imply the reality 
condition $D=D^\dag$ and $\rho^{ij}=\bar\rho^{ij}$ where 
$\bar\rho_{ij}=(\rho^{ij})^\dag$. Condition \eqref{2} yields
\begin{equation}
\sud{\sud H i \alpha} j {\dot\alpha} = i \epsilon^{ij} 
h_{\alpha\dot\alpha}.
\label{c3}
\end{equation}

Higher components of the superfield constraint \eqref{1}--\eqref{4} 
imply further conditions on the fields $f_{\alpha\beta}$, 
$h_{\alpha\dot\alpha}$, and $D$. One way to realize these conditions is 
to note that, from equations \eqref{3.32}, \eqref{3.30}, and 
\eqref{3.29},
\begin{align}
\mathscript{F}_{ab} &= F_{ab}| = - \tfrac18 \du{\bar\sigma} a 
{\dot\alpha\alpha} \du{\bar\sigma} b {\dot\beta\beta} \left( 
\epsilon_{\dot\alpha\dot\beta} f_{\alpha\beta} + \epsilon_{\alpha\beta} 
\bar f_{\dot\alpha\dot\beta} \right),\label{b2} \\
\left. F_{a\bar z} \right| &= \tfrac12 h_a, \label{F1} \\
F_{z\bar z}| &= 2 i D, \label{F2}
\end{align}
where $\bar f_{\dot\alpha\dot\beta} = (f_{\alpha\beta})^\dag$ and $h_a 
= -\tfrac12 \du \sigma a {\alpha\dot\alpha} h_{\alpha\dot\alpha}$. The 
equations \eqref{3.29}--\eqref{4} are the general solution of the 
Bianchi identities subject to our constraints. It was shown in 
\cite{PLB-392-85}, that three of these Bianchi identities imply that 
$\mathscript F_{ab}$ is the field strength of a gauge field 
$V_a$
\begin{equation}
\mathscript F_{ab}=\partial_a V_b - \partial_b V_a.
\label{b3}
\end{equation}
and that 
\begin{align}
\partial_{[a} h^R_{b]} &= - \tfrac12 (\partial_z + \partial_{\bar z}) 
\mathscript F_{ab}, \\
\partial_{[a} h^I_{b]} &= - \tfrac12 i (\partial_z-\partial_{\bar z}) 
\mathscript F_{ab}, \label{h1} \\
\partial_a D &= - \tfrac14 i \left\{ (\partial_z-\partial_{\bar z}) 
h_a^R  + i (\partial_z + \partial_{\bar z}) h_a^I \right\},
\label{temp11}
\end{align}
where $h_a = h_a^R + i h_a^I$.

This is as far as we can go using solely the constraints \eqref{con-1}. 
So far we have the following component fields: an $\SU(2)$ doublet of 
spinors $\lambda^i_{\alpha}$, a real gauge field $V_a$, a complex 
vector field $h_a$, a real scalar $D$ and a real $\SU(2)$ triplet of 
scalars $\rho^{ij}$. We would like to impose further constraints to 
reduce the number of fields to an irreducible multiplet. First consider
the fields $D$ and $\rho^{ij}$? We will find that $D$ and 
$\rho^{ij}$ play the role of auxiliary fields. Note that the usual 
$N=2$ gauge multiplet has a triplet of auxiliary fields. Hence, if we 
want to derive the usual gauge multiplet we are led to set $D$ to zero 
by promoting the constraint \eqref{3} to the stronger one
\begin{equation}
\epsilon^{\alpha\beta} D_\alpha^{[i} W_\beta^{j]} = 0, \qquad  
\epsilon^{\dot\alpha\dot\beta} \bar D_{\dot\alpha}^{[i} \bar 
W_{\dot\beta}^{j]} = 0.
\label{f1}
\end{equation}
We will discuss this constraint, in the non-Abelian context, later.
On the other hand, if we want to derive the vector-tensor multiplet, 
which has a single auxiliary field, we must eliminate $\rho^{ij}$. This 
can be achieved by promoting condition \eqref{4} to the stronger 
condition
\begin{equation}
\epsilon^{\alpha\beta} D_\alpha^{(i} W_\beta^{j)} = 0, \qquad 
\epsilon^{\dot\alpha\dot\beta} \bar D_{\dot\alpha}^{(i} \bar 
W_{\dot\beta}^{j)} = 0.
\label{4'}
\end{equation}
Henceforth, we add condition \eqref{4'} as a further constraint on the 
theory. Now let us proceed using conditions \eqref{1}--\eqref{3} as 
well as the stronger version of \eqref{4}, constraint \eqref{4'}. Apart 
from eliminating $\rho^{ij}$, it can be shown that the consequences of 
imposing \eqref{4'} are
\begin{equation}
\begin{split}
\partial^a h_a^R &= i (\partial_z - \partial_{\bar z}) D, \\
\partial^a h_a^I &= - (\partial_z + \partial_{\bar z}) D, \\
\partial^b \mathscript F_{ab} &= -\tfrac14 \left\{ 
(\partial_z+\partial_{\bar z}) h_a^R + i (\partial_z - \partial_{\bar 
z}) h_a^I \right\}.
\end{split}
\label{h2}
\end{equation}
We would like to render the real and 
imaginary components of $h_a$ as field strengths. 
To do this, we must impose yet another, and final, constraint.
We take a reality condition on the central charge
\begin{equation}
\partial_z W^i_\alpha = \partial_{\bar z} W^i_\alpha.
\label{realz}
\end{equation}
This means that superfield $W^i_\alpha$, and hence all the curvature 
tensor components, depend on the two central charges $z$ and $\bar z$ 
only through the combination $z + \bar z$ with no dependence on $z-\bar 
z$. Substituting from constraint \eqref{realz} into \eqref{h1} and 
\eqref{h2} immediately gives the following constraints on the real and 
imaginary components of $h_a$
\begin{equation}
\partial^a h_a^R = 0, \qquad \partial_{[a} h_{b]}^I = 0.
\label{t5}
\end{equation}
They assert 
that $h_a^R$ and $h_a^I$ are field strengths of an anti-symmetric 
tensor and a scalar field respectively. That is
\begin{equation}
\begin{split}
h_a^R &= \tfrac13 \epsilon_{abcd} H^{bcd} = \tfrac13 \epsilon_{abcd} 
\partial^b B^{cd}, \\
h_a^I &= 2 \partial_a \phi.
\end{split}
\label{b1}
\end{equation}

In conclusion, we find that the component fields in $W^i_\alpha$ are 
exactly those of the vector-tensor multiplet, namely 
$(\lambda^i_\alpha, \phi, B_{ab}, V_a, D)$. These fields are actually 
functions of the 
central charge coordinates as well as the spacetime coordinates. 
However, by virtue of constraint \eqref{realz}, they depend on the 
central charge coordinates only via $z+\bar z$. Let us introduce a real 
central charge coordinate $\cc=z+\bar z$. We now show that the above 
conditions completely determine 
their dependence on $\cc$ in terms of their lowest component in their 
central charge expansion. To see this, let us summarize the relevant 
conditions in \eqref{h1}, \eqref{temp11}, and \eqref{h2} derived on the 
bosonic fields. They are
\begin{equation}
\begin{split}
\partial_\cc D &= - \tfrac12 \partial^a h_a^I, \\
\partial_\cc h_a^R &= -2 \partial^b \mathscript F_{ab}, \\
\partial_\cc h_a^I &= 2 \partial_a D, \\
\partial_\cc \mathscript F_{ab} &= - \partial_{[a} h_{b]}^R.
\end{split}
\label{t4}
\end{equation}
Equations \eqref{t4} constitute a system of first-order ``differential 
equations'' for the $s$ dependence of the bosonic fields. The general 
solution is completely determined in terms of the ``initial 
conditions''; that is, the values of the fields at $s=0$, say, 
$D(x^m)$, $h_a(x^m)$, and $\mathscript F_{ab}(x^m)$. Thus in a Taylor 
series expansion in $\cc$ each term in the series gets related to 
lower-order terms, so that for instance,
\begin{equation}
D(x^m,\cc) = D(x^m) - \tfrac12 \partial^a h_a^I(x^m) \cc - \tfrac12 
\Box D(x^m) \cc^2 + \cdots,
\end{equation}
with similar formulas for $h_a^R$, $h_a^I$, and $\mathscript F_{ab}$. 
Furthermore, in \cite{PLB-392-85} we showed that
\begin{equation}
\partial_\cc \lambda^i_\alpha = \ud\sigma a {\alpha\dot\alpha} 
\partial_a \bar\lambda^{\dot\alpha i},
\label{ccl}
\end{equation}
where $\bar\lambda_{\dot\alpha i} = (\lambda^i_{\alpha})^\dag$. 
Equation \eqref{ccl} is the relation for the fermionic field 
$\lambda_i^\alpha$ corresponding to equations \eqref{t4} for the 
bosonic fields. It fixes the expansion of $\lambda^i_\alpha$ in the 
central charge $\cc$, leaving only the lowest component 
$\lambda^i_\alpha(x^m)$ arbitrary.

We now compute the supersymmetry transformation of 
the different component fields. To do this we have to act on 
$W^i_\alpha$ with
\begin{equation}
\delta_\xi = \xi^\alpha_i Q_\alpha^i + \bar\xi_{\dot\alpha}^i \bar 
Q_i^{\dot\alpha}.
\label{c4}
\end{equation}
We find
\begin{equation}
\begin{split}
\delta_\xi D & = - \tfrac12 i \left( \xi^\alpha_i \ud\sigma a 
{\alpha\dot\alpha} \partial_a \bar\lambda^{\dot\alpha i} + 
\bar\xi^{\dot\alpha}_i \ud\sigma a {\alpha\dot\alpha} \partial_a 
\lambda^{\alpha i} \right), \\
\delta_\xi \phi &= \tfrac12 i \left( \xi_{\alpha i} \lambda^{\alpha i} 
- \bar\xi_{\dot\alpha i} \bar\lambda^{\dot\alpha i} \right), \\
\delta_\xi B_{cd} &= \tfrac16 \epsilon_{abcd} \ud \sigma {ab} 
{(\alpha\beta)} \xi^\alpha_i \lambda^{\beta j} + \tfrac16 
\epsilon_{abcd} \ud {\bar\sigma} {ba} {(\dot\alpha\dot\beta)} 
\bar\xi^{\dot\alpha}_i \bar\lambda^{\dot\beta i}, \\
\delta_\xi V_a & = - \tfrac12 \xi_{\alpha i} \du {\bar\sigma} a 
{\dot\alpha\alpha} \bar \lambda^k_{\dot\alpha} + \tfrac12 
\bar\xi_{\dot\alpha i} \du{\bar\sigma} a {\dot\alpha\alpha} 
\lambda^k_\alpha, \\
\delta_\xi \lambda^i_\alpha & = 2 i \xi^{\beta i} 
\epsilon^{\dot\alpha\dot\beta} \ud\sigma a {\alpha\dot\alpha} \ud\sigma 
b {\beta\dot\beta} \partial_{[a} V_{b]} 
+ 2 \xi_\alpha^i D 
+ \tfrac{i}{3} \bar\xi^{\dot\alpha i} \ud\sigma a {\alpha\dot\alpha} 
\epsilon_{abcd} \partial^b B^{cd} - 2 \ud \sigma a {\alpha\dot\alpha} 
\bar\xi^{\dot\alpha} \partial_a \phi.
\end{split}
\label{susy}
\end{equation}
The fields in these supersymmetry transformations are 
formally functions of both $x^m$ and $\cc$, but \eqref{susy}
 also express the variations of the lowest-order 
independent fields, which are functions of $x^m$ only.

To get the central charge transformations we have to act with 
$\delta_\omega= \omega\partial_\cc$ on $W^i_\alpha$. It is 
straightforward to get the following set of transformations
\begin{equation}
\begin{split}
\delta_\omega D & = - \omega \Box \phi, \\
\delta_\omega \phi &= \omega D, \\
\delta_\omega B^{cd} &= 3 \omega \epsilon^{abcd} \partial_{a} V_b,\\
\delta_\omega V_a & = - \tfrac{1}{6} \omega \epsilon_{abcd} \partial^b 
B^{cd}, \\ 
\delta_\omega \lambda^i_\alpha & = \omega \ud\sigma a 
{\alpha\dot\alpha} \partial_a \bar\lambda^{\dot\alpha i}.
\end{split}
\label{cc}
\end{equation}

Expressions \eqref{susy} and \eqref{cc} reproduce the supersymmetry and 
central charge transformations of the vector-tensor multiplet given in 
the component field calculations of 
\cite{PLB-92-123,NPB-173-127,NPB-451-53}. 

We now give the superfield action for the pure vector-tensor 
multiplet. In \cite{PLB-392-85} it was found that the appropriate action is
\begin{align}
S &= \int d^4x \left\{ - \tfrac1{192} D_i^\alpha D_{\alpha j} W^{\beta 
i} W^j_\beta + \tfrac1{192}\bar D_{\dot\alpha i} \bar D^{\dot\alpha}_j 
W^{\beta i} W^j_\beta + \text{h.c.} \right\} \notag \\
&= \int d^4x \left\{ 
- \tfrac14 \mathscript F_{ab} \mathscript F^{ab} - \tfrac12 \partial_a 
\phi \partial^a \phi 
- \tfrac{1}{12} \partial_a B_{bc} \partial^a B^{bc} 
+ \tfrac12 D^2 - \tfrac14 i \lambda^{\alpha i} \ud\sigma a 
{\alpha\dot\alpha} \partial_a \bar\lambda^{\dot\alpha}_i \right\}.
\label{zz}
\end{align}
This represents a central charge generalization of the type of 
superactions considered in \cite{NPB-191-445}. This completes our brief
discussion of the gauge superfield formulation of the vector-tensor 
multiplet.

\section{Integrability and the Superfield W}

The results of the previous section are sufficient to describe the vector-
tensor multiplet. However, there is an, as yet, undiscussed property 
of this theory which gives considerable insight into the nature of this 
multiplet. The constraints on superfield $W^{i}_{\alpha}$ are given in 
\eqref{1}-\eqref{3}, \eqref{4'} and \eqref{realz}. These constraints are all 
first order differential conditions. Further differentiation of these 
constraints leads to a number of integrability conditions. These are 
sufficient to allow one to completely solve for $W^{i}_{\alpha}$ in terms of 
a real scalar superfield. This scalar superfield is subject to predominantly 
second order constraints. Specifically, we find that
\begin{equation}
W^i_\alpha = i D^i_\alpha W,
\label{w}
\end{equation}
where scalar superfield $W$ is real, that is $W^\dagger=W$. 
The form of \eqref{w} follows immediately from the $W^{i}_{\alpha}$
constraint \eqref{1} and the fact that
\begin{equation}
D^{(i}_{(\alpha} D^{j)}_{\beta)} =0
\end{equation}
The
superfield $W$ is subject to a set of constraints that can be
obtained by substituting \eqref{w} into \eqref{2}, \eqref{3},
\eqref{4'}, and \eqref{realz}. 
Substituting expression \eqref{w} into $W^i_\alpha$ constraints \eqref{2}
 and
\eqref{4'} implies that $W$ must satisfy the conditions
\begin{equation}
D^{(i}_\alpha \bar D^{j)}_{\dot\beta} W = 0, \qquad 
\bar D^{(i}_{\dot\alpha} D^{j)}_\beta W = 0
\label{b5}
\end{equation}
and
\begin{equation}
\epsilon^{\alpha\beta}D^{(i}_{\alpha} D^{j)}_{\beta} W = 0, \qquad 
\epsilon^{\dot\alpha \dot\beta}\bar D^{(i}_{\dot\alpha} 
\bar D^{j)}_{\dot\beta} W = 0
\label{b6}
\end{equation}
respectively. Finally, using the fact that
\begin{equation}
D^{[i}_{[\alpha}D^{j]}_{\beta]}=-i\epsilon^{ij}\epsilon_{\alpha\beta}
\partial_{z}
\end{equation}
we find from $W^i_\alpha$ constraint \eqref{3} that
\begin{equation}
\partial_z W = \partial_{\bar z} W.
\label{c5}
\end{equation}
The appearance of $W$ in this gauge formulation is, at first sight,
somewhat surprising since all the curvature tensor components are given in 
terms of $W^i_\alpha$ and its derivatives. However, as we will show in a 
forthcoming paper \cite{inpre}, there exists a one-parameter class of 
constraints, all of which describe the vector-tensor multiplet. The set of 
constraints that we use in this paper is a single instance of this class of 
constraints. In all but this particular instance, the scalar superfield $W$ 
appears explicitly and naturally as a component of the curvature tensor. 

Note that since $W^i_\alpha$ is a gauge-invariant object, so is $W$. Since 
$W^i_\alpha$ and, hence, all other physically relevant superfields can be 
obtained from $W$ by differentiation, it is clear
that $W$ plays a fundamental role in the description of the
vector tensor-multiplet. In particular, $W$ should contain all the 
component fields of the vector-tensor multiplet. Indeed it does, but in a 
somewhat non-trivial way. To see this, let us postulate the following
expansion of $W$ in the anti-commuting coordinates
\begin{equation}
W = 2\phi - i \theta_i^\alpha \lambda_\alpha^i + i
\bar\theta^{\dot\alpha}_i \bar\lambda^i_{\dot\alpha} + \theta^\alpha_i
\bar\theta^{\dot\alpha i} h^{R}_{\alpha\dot\alpha} + \tfrac12
\theta^\alpha_i \theta^{\beta i} f_{\alpha\beta} - \tfrac12
{\bar\theta}_{i}^{\dot\alpha} {\bar\theta}^{\dot\beta i}{\bar
f}_{\dot\alpha\dot\beta} + \mathscript{O}(\theta^3).
\label{b4}
\end{equation}
We remind the reader that the component fields 
satisfy \eqref{b2}, \eqref{b3}
and \eqref{b1}. Substituting this into relation \eqref{w}, we find that it
correctly reproduces the component expansion \eqref{c1}-\eqref{c3} of 
$W^i_\alpha$ 
as long
as one can identify the derivative term $\partial_{s}\phi$ with the 
auxiliary
field $D$. This identification can be justified using 
the central charge conditions \eqref{t4}.
Specifically, combining the definition of $\phi$ in \eqref{b1} with 
the equations \eqref{t4}, we find that
\begin{equation}
\phi(x^{m},s)=\phi(x^{m})+D(x^{m})s-\tfrac{1}{2}\Box\phi(x^{m})s^{2}+...
\end{equation}
and therefore
\begin{equation}
D=\partial_{s}\phi|_{s=0}
\end{equation}
Conditions \eqref{t4} further imply that all other coefficients in the $s$ 
expansion of $W$ are determined in terms of $\lambda^{i}_{\alpha},
\phi, B_{ab}, V_{a}$ and $D$.
We conclude that \eqref{b4} is indeed the correct component expansion
of scalar superfield $W$ and that it contains all of the component fields
$\lambda^{i}_{\alpha}, \phi, B_{ab}, V_{a}$ and $D$. It is interesting to 
note that the component
fields  $\lambda^{i}_{\alpha}, \phi, B_{ab}$ and $V_{a}$ all occur at zeroth
order in the central charge, whereas 
auxiliary field $D$ appears at first order
in the Taylor expansion of $\phi$ in the central charge $s$.
This is similar to the appearance of the auxiliary fields $\phi_i$ in the
Fayet-Sohnius multiplet. In that superfield, the lowest order
piece gives the dynamically propagating fields $A_i$, while the first
order term in the $s$-expansion gives the auxiliary fields.
In the above discussion, we derived the component expansion of $W$ making
use of our previous results for $W^{i}_{\alpha}$. It is important to 
emphasize, however, that this is not necessary. The conditions that 
superfield $W$ be real and satisfy constraints \eqref{b5}, \eqref{b6} 
and \eqref{c5} are sufficient to completely determine the component 
structure. Furthermore, the supersymmetry transformations \eqref{susy} and 
the central charge transformations \eqref{cc} can be determined directly by 
acting on $W$ with \eqref{c4} and $\delta_{\omega}=\omega\partial_{s}$.

We close by emphasizing that the vector-tensor multiplet is formulated in 
this section as a gauge multiplet, where the component gauge field $V_{a}$ 
is introduced directly in the one-form connection and the appearance of the 
antisymmetric tensor component field $B_{ab}$ is a consequence of applying 
the appropriate constraints. Note, however, that the $N=2$ superfield $W$ 
bears a close resemblance to the real linear multiplet superfield of $N=1$ 
supersymmetry. For $N=1$ theories, a geometric superspace formulation of the 
linear multiplet in terms of a two-form multiplet was constructed in 
\cite{NPB-292-181}. This suggests that the $N=2$ vector-tensor multiplet has 
a similar formulation in terms of an $N=2$ two-form multiplet as
well. This will be the subject of Section~5.

\section{Vector-Tensor Multiplet: Two-Form Formulation}

One begins by introducing a real two-form superfield in superspace with
central charge.
\begin{equation}
B=\tfrac{1}{2}e^{B}e^{A}B_{AB}
\end{equation}
where $e^{A}$ denotes the frame of flat superspace. The invariant field
strength is the three-form
\begin{equation}
H=dB
\end{equation}
Its component superfields, defined by
\begin{equation}
H=\frac{1}{3!}e^{C}e^{B}e^{A}H_{ABC}
\end{equation}
are then subject to a set of constraints. The first of these stipulates
that any completely spinorial component of $H$ must vanish. That is
\begin{equation}
H^{ijk}_{\underline{\alpha} \underline{\beta}
\underline{\gamma} } = 0.
\end{equation}
The second constraint is given by
\begin{equation}
H^{ij}_{\alpha\beta a} = 0, \qquad H^{ij}_{\dot\alpha \dot\beta a}
= 0.
\end{equation}
The third, and final, constraint is that
\begin{equation}
H^{ij}_{\alpha\dot\beta a} = H^{ji}_{\dot\beta\alpha a} =
2i \epsilon^{ij} \sigma_{a\alpha\dot\beta} L
\end{equation}
where $L$ is a real superfield. We now solve the Bianchi identities, 
$dH=0$, subject to these constraints. We find that all component superfields 
of $H$ vanish except for the following
\begin{equation}
H^{ij}_{\alpha\beta \bar z} = -4i \epsilon^{ij}
\epsilon_{\alpha\beta} L, \qquad H^{ij}_{\dot\alpha\dot\beta z} 
= - 4 i \epsilon^{ij}\epsilon_{\dot\alpha\dot\beta} L,
\end{equation}
\begin{equation}
H^k_{\gamma \bar z z} = 2 D^k_\gamma L,  \qquad H^k_{\dot\gamma z \bar z} = 
2 D^k_{\dot\gamma} L,
\end{equation}
\begin{equation}
H^i_{\alpha \bar z a} = -2 \epsilon_{\alpha\beta}
\du {\bar\sigma} a {\dot\gamma \beta} \bar D^i_{\dot\gamma} 
L, \qquad H^i_{\dot\alpha z a} = 2 \epsilon_{\dot\alpha\dot\beta}
\du {\bar\sigma} a {\dot\beta \gamma} D^i_\gamma L,
\end{equation}
\begin{equation}
H^k_{\gamma a b} = -2 {\sigma_{ab\gamma}}^\alpha
D^k_\alpha L, \qquad H^k_{\dot\gamma a b} = 2 \dud {\bar\sigma} {ab} 
{\dot\alpha} {\dot\gamma} \bar D^k_{\dot\alpha} L,
\end{equation}
\begin{equation}
H_{zab} = - \frac{i}{2} \epsilon_{ij} \epsilon^{\alpha\gamma} 
\du \sigma {ab\gamma} \beta D^i_\alpha D^j_\beta L, \qquad H_{\bar z ab} = 
\frac{i}{2} \epsilon_{ij} \epsilon^{\dot\alpha\dot\gamma} \dud {\bar\sigma} 
{ab} {\dot\beta} {\dot\gamma} \bar D_{\dot\alpha}^i \bar D_{\dot\beta}^j L,
\end{equation}
\begin{equation}
H_{az\bar z} = - \frac{i}{4} \epsilon_{ij} \du {\bar\sigma} a 
{\dot\alpha\alpha} [D^i_\alpha, \bar D^j_{\dot\alpha}] L,
\qquad H_{abc} = - \tfrac18 \epsilon_{abcd} \bar\sigma^{d \dot\beta \alpha}
\epsilon_{ij} [D^i_\alpha, \bar D^j_{\dot\beta}] 
L.
\end{equation}
Furthermore, superfield $L$ is required to satisfy
\begin{equation}
D^{(i}_\alpha \bar D^{j)}_{\dot\beta} L = 0
\label{z3}
\end{equation}
as well as 
\begin{equation}
\epsilon^{\alpha\beta} D^{(i}_{\alpha} D^{j)}_{\beta} L =
0, \qquad \epsilon^{\dot\alpha\dot\beta} \bar D^{(i}_{\dot\alpha}
\bar D^{j)}_{\dot\beta} L = 0
\label{z4}
\end{equation}
It is important to note that these conditions on $L$ are the
same as constraints \eqref{b5} and \eqref{b6} on superfield $W$. 
If we impose one 
additional constraint on $L$, namely that
\begin{equation}
\partial_{z}L=\partial_{\bar z}L
\end{equation}
then the constraints on $L$ are the exactly same as those on $W$.  
As discussed at the end of the previous section, the conditions that 
superfield $L$ be real and satisfy constraints \eqref{b5}, \eqref{b6} 
and \eqref{c5} are sufficient
to completely determine its component field structure. Since the constraints
are identical to those on $W$, it follows that superfield $L$ describes the 
vector-tensor multiplet with component fields $\lambda^{i}_{\alpha}, \phi, 
B_{ab}, V_{a}$ and $D$.
Furthermore, the 
supersymmetry transformations \eqref{susy} and the central charge 
transformations \eqref{cc}
can be determined directly by acting on $L$ with \eqref{c4} and 
$\delta_{\omega}=\omega\partial_{s}$. That is, we can identify
\begin{equation}
L=W
\end{equation}
We conclude that the vector-tensor multiplet can be equivalently described
by introducing the two-form component field $B_{ab}$ directly through 
a two-form superfield and deriving the gauge component field $V_{a}$ by 
applying the appropriate constraints. 

\section{Non-Abelian Superauge Multiplets and Chern-Simons Forms}

Having presented both a supergauge field formulation and a super three-form
formulation of the vector-tensor multiplet, we now return to the generic
supergauge theory of Section 3. We modify that theory so as to describes 
usual $N=2$ supergauge multiplets associated with a non-Abelian gauge group.
This modification is very straightforward. We begin by introducing a Lie
algebra valued Hermitian connection 
$\mathscript{A}=dz^{M}\mathscript{A}=e^{A}
\mathscript{A}_{A}$. The hermiticity of the connection then implies
\begin{equation}
\mathscript{A} = dx^a \mathscript{A}_a + d\theta^\alpha_i 
\mathscript{A}^i_\alpha + 
d\bar\theta^i_{\dot\alpha} 
\bar \mathscript{A}_i^{\dot\alpha} + dz \mathscript{A}_z + d\bar z 
\mathscript{A}
_{\bar z},
\end{equation}
where $\mathscript{A}_a$ is real, $\bar \mathscript{A}^{\dot\alpha}_i = 
(\mathscript{A}
^{\alpha i})^\dag$, and 
$\mathscript{A}_{\bar z} = \mathscript{A}_z^\dag$. The curvature two-form is 
defined as
\begin{equation}
\mathscript{F} = d \mathscript{A}+\mathscript{A} \mathscript{A} = \tfrac12 
e^B e^A 
\mathscript{F}_{AB}.
\end{equation}
The curvature tensor $\mathscript{F}$ is subject to the Bianchi identities 
$\mathscript{D}\mathscript{F}=0$.
As in the Abelian case, we will adopt a set of constraints 
which set the pure spinorial part of the curvature tensor to zero. That 
is
\begin{equation}
\sud{\sud {\mathscript{F}} i \alpha} j \beta = 
\sud{\sud {\mathscript{F}} i \alpha} j {\dot\beta} = 
\sud{\sud {\mathscript{F}} i {\dot\alpha}} j {\dot\beta} = 0.
\label{e1}
\end{equation}
Before imposing further conditions we would like to explore the 
consequences of \eqref{e1}. To do so we must solve the Bianchi 
identities subject to these constraints. The result is that all the 
components of the curvature tensor $\mathscript{F}_{AB}$ are determined in 
terms of 
a single superfield $\mathscript{F}^i_{\alpha\bar z}$ and its hermitian 
conjugate 
$\mathscript{F}_{\dot\alpha i z} = (\mathscript{F}^i_{\alpha \bar z})^\dag$. 
Henceforth, we 
denote these superfields by $\mathscript{W}^i_\alpha$ and $\bar 
\mathscript{W}_{\dot\alpha i}$. 
In particular $\mathscript{F}^i_{\alpha z} = 0$ and
\begin{align} 
\mathscript{F}_{ab} &= -\tfrac1{16} i \epsilon_ {ij} 
\du{\bar\sigma}a{\dot\alpha\alpha} \du{\bar\sigma}b{\dot\beta\beta} 
\left( \epsilon_{\dot\alpha\dot\beta} 
\mathscript{D}^j_\beta\mathscript{W}^i_\alpha + 
\epsilon_{\alpha\beta} \bar \mathscript{D}^j_{\dot\beta} \bar 
\mathscript{W}^i_{\dot\alpha} 
\right). \label{e2} \\
\mathscript{F}^i_{\dot\alpha a} &= - \tfrac12 \epsilon_{\dot\alpha\dot\beta} 
\du{\bar\sigma}a{\dot\beta\alpha} \mathscript{W}^i_\alpha, \label{ee2} \\ 
\mathscript{F}_{a\bar z} &= - \tfrac18 i \epsilon_{ij} 
\du{\bar\sigma}a{\dot\beta\alpha} \bar \mathscript{D}^j_{\dot\beta} 
\mathscript{W}^i_\alpha, 
\label{e3} \\
\mathscript{F}_{z\bar z}  &= \tfrac14 i \epsilon_{ij} \epsilon^{\alpha\beta} 
\mathscript{D}_\alpha^i \mathscript{W}_\beta^j. \label{e4}
\end{align}
Furthermore, $\mathscript{W}^i_\alpha$ is constrained to satisfy 
\begin{align}
&\mathscript{D}^{(j}_{(\beta} \mathscript{W}^{i)}_{\alpha)} = 0, \qquad 
\bar \mathscript{D}^{(j}_{(\dot\beta} 
\bar \mathscript{W}^{i)}_{\dot\alpha)} = 0, \label{e5} \\
& \bar \mathscript{D}^{(j}_{\dot\beta} \mathscript{W}^{i)}_{\alpha} = 0, 
\qquad 
\mathscript{D}^{(j}_\beta \bar 
\mathscript{W}^{i)}_{\dot\alpha} = 0, \label{e6} \\
& \epsilon^{\alpha\beta} \mathscript{D}_\alpha^{[i} 
\mathscript{W}_\beta^{j]} 
= - \epsilon^{\dot\alpha\dot\beta} \bar \mathscript{D}_{\dot\alpha}^{[i} 
\bar 
\mathscript{W}_{\dot\beta}^{j]}, \label{e7} \\
& \epsilon^{\alpha\beta} \mathscript{D}_\alpha^{(i} 
\mathscript{W}_\beta^{j)} 
= \epsilon^{\dot\alpha\dot\beta} \bar \mathscript{D}_{\dot\alpha}^{(i} \bar 
\mathscript{W}_{\dot\beta}^{j)}.
\label{e9}
\end{align}
Now, as discussed in Section 3, it is necessary to 
impose additional constraints to further reduce the number of fields to
an irreducible supermultiplet. In order to obtain the usual gauge multiplet 
in the Abelian case, one introduces constraint \eqref{f1}. The non-Abelian
generalization of this is straightforward and is obtained by promoting
constraint \eqref{e7} to the stronger one
\begin{equation}
\epsilon^{\alpha\beta} \mathscript{D}_\alpha^{[i} \mathscript{W}_\beta^{j]} 
= 0, \qquad  
\epsilon^{\dot\alpha\dot\beta} \bar \mathscript{D}_{\dot\alpha}^{[i} \bar 
\mathscript{W}_{\dot\beta}^{j]}  = 0.
\label{x2}
\end{equation}
Note that this additional constraint is equivalent to setting
\begin{equation}
\mathscript{F}_{z \bar z}=0.
\label{x3}
\end{equation}
Furthermore, it is necessary to add a reality condition on the central
charge given by
\begin{equation}
\mathscript{D}_{z}\mathscript{W}^{i}_{\alpha}=\mathscript{D}_{\bar 
z}\mathscript{W}^{i}_{\alpha}, \qquad
\mathscript{D}_{z}\bar\mathscript{W}^{i}_{\dot\alpha}=\mathscript{D}_{\bar 
z}.
\bar\mathscript{W}^{i}_{\dot\alpha}
\label{x1}
\end{equation}
Constraints \eqref{e1}, \eqref{x2} (or, equivalently, \eqref{x3}) and 
\eqref{x1} are sufficient to completely describe the non-Abelian
supergauge multiplet. Now, as in the case of the vector-tensor
multiplet, these constraints are such as to allow us to completely solve for
$\mathscript{W}^{i}_{\alpha}$ and $\bar\mathscript{W}^{i}_{\dot\alpha}$ in 
terms of two complex scalar superfields $\mathscript{W}$ and 
$\bar\mathscript{W}$. Specifically, we find that
\begin{equation}
\mathscript{W}^{i}_{\alpha}=i\mathscript{D}^{i}_{\alpha}\mathscript{W}, 
\qquad
\bar\mathscript{W}^{i}_{\dot\alpha}= - i \bar\mathscript{D}^{i}_{\dot\alpha}
\bar\mathscript{W},
\label{x4}
\end{equation}
where scalar superfield $\mathscript{W}$ is complex and $\bar\mathscript{W}=
\mathscript{W}^{\dag}$. The form of \eqref{x4} follows immediately from the
constraint \eqref{e5} and the fact that
\begin{equation}
\mathscript{D}^{(i}_{(\alpha}\mathscript{D}^{j)}_{\beta)}=0.
\label{x5}
\end{equation}
The superfields  $\mathscript{W}$ and $\bar\mathscript{W}$ are subject to a 
set of constraints that can be obtained by substituting \eqref{x4} 
into \eqref{e6}, \eqref{e9}, \eqref{x2} and \eqref{x1}. Substituting 
expression \eqref{x4} into constraints \eqref{e6} and \eqref{e9} implies 
that $\mathscript{W}$ and $\bar\mathscript{W}$ must satisfy the conditions
\begin{equation}
\bar\mathscript{D}^{(i}_{\dot\alpha}\mathscript{D}^{j)}_{\beta}\mathscript{W
}=0,
\qquad
\mathscript{D}^{(i}_{\alpha} \bar\mathscript{D}^{j)}_{\dot\beta} 
\bar\mathscript{W}=0
\label{x6}
\end{equation}
and
\begin{equation}
\epsilon^{\alpha 
\beta}\mathscript{D}^{(i}_{\alpha}\mathscript{D}^{j)}_{\beta}
\mathscript{W}=\epsilon^{\dot\alpha \dot\beta}
\bar\mathscript{D}^{(i}_{\dot\alpha}\bar\mathscript{D}^{j)}_{\dot\beta}
\bar\mathscript{W}
\label{x7}
\end{equation}
respectively. Using the fact that 
\begin{equation}
\mathscript{D}^{[i}_{[\alpha}\mathscript{D}^{j]}_{\beta]}=-i\epsilon^{ij}
\epsilon_{\alpha \beta}\mathscript{D}_{z},
\end{equation}
we find from constraint \eqref{x2} that
\begin{equation}
\mathscript{D}_{z}\mathscript{W}=0, \qquad
\mathscript{D}_{\bar z}\bar\mathscript{W}=0.
\label{y1}
\end{equation}
Combining this result with constraint \eqref{x1} we have, finally, that
\begin{equation}
\mathscript{D}_{z}\mathscript{W}=\mathscript{D}_{\bar z}\mathscript{W}=0,
\end{equation}
\begin{equation}
\mathscript{D}_{z}\bar\mathscript{W}=\mathscript{D}_{\bar 
z}\bar\mathscript{W}=0.
\label{y2}
\end{equation}
As will be discussed elsewhere, it is possible, using the
redundancy in the solution of \eqref{x4}, to choose $\mathscript{W}$ 
and $\bar\mathscript{W}$ to satisfy 
$\bar\mathscript{D}^{i}_{\dot\alpha}\mathscript{W}=
\mathscript{D}^{i}_{\alpha}\bar\mathscript{W}=0$. We do this henceforth.

The Chern-Simons form associated with this non-Abelian supergauge
theory is defined to be
\begin{equation}
\mathscript{Q}=\tr(\mathscript{A}\mathscript{F}-\tfrac{1}{3}\mathscript{A} 
\mathscript{A}\mathscript{A})
\label{g1}
\end{equation}
The Chern-Simons term can be coupled to the super two-form $B$ discussed
in the previous section by modifying the three-form superfield strength
$H$ to become
\begin{equation}
\mathscript{H}=dB+\kappa \mathscript{Q}
\label{g2}
\end{equation}
where $\kappa$ is an arbitrary real coupling parameter.
Its component superfields, defined by
\begin{equation}
\mathscript{H}=\frac{1}{3!}e^{C}e^{B}e^{A}\mathscript{H}_{ABC}
\end{equation}
are then subject to the constraints
\begin{equation}
\mathscript{H}^{ijk}_{\underline{\alpha} \underline{\beta}
\underline{\gamma} } = 0.
\end{equation}
\begin{equation}
\mathscript{H}^{ij}_{\alpha\beta a} = 0, 
\qquad \mathscript{H}^{ij}_{\dot\alpha \dot\beta a}
= 0.
\end{equation}
\begin{equation}
\mathscript{H}^{ij}_{\alpha\dot\beta a} = 
\mathscript{H}^{ji}_{\dot\beta\alpha a} =
2i \epsilon^{ij} \sigma_{a\alpha\dot\beta} \mathscript{L}
\end{equation}
where $\mathscript{L}$ is a real superfield. The addition of the 
Chern-Simons form modifies the Bianchi identities to
\begin{equation}
d\mathscript{H}=\kappa \tr\mathscript{F}\mathscript{F}
\end{equation}
We now solve these Bianchi identities subject to the above constraints. We 
find that
\begin{equation}
\mathscript{H}^{ij}_{\alpha\beta \bar z} = -4i \epsilon^{ij}
\epsilon_{\alpha\beta} \mathscript{L}, 
\qquad \mathscript{H}^{ij}_{\dot\alpha\dot\beta z} = 
- 4 i \epsilon^{ij}\epsilon_{\dot\alpha\dot\beta} \mathscript{L}.
\end{equation}
\begin{equation}
\mathscript{H}^k_{\gamma \bar z z} = 2 D^k_\gamma \mathscript{L},  
\qquad  \mathscript{H}^k_{\dot\gamma z \bar z}
= 2 D^k_{\dot\gamma} \mathscript{L}.
\end{equation}
\begin{equation}
\mathscript{H}^i_{\alpha \bar z a} = -2 \epsilon_{\alpha\beta}
{\bar\sigma}_a^{\dot\gamma\beta} \bar D^i_{\dot\gamma} 
\mathscript{L}, \qquad \mathscript{H}^i_{\dot\alpha z a} = 2 \epsilon_
{\dot\alpha\dot\beta}
{\bar\sigma}_a^{\dot\beta\gamma} D^i_\gamma \mathscript{L}.
\end{equation}
\begin{equation}
\mathscript{H}^k_{\gamma a b} = - 2 {\sigma_{ab\gamma}}^\alpha
D^k_\alpha \mathscript{L}, \qquad \mathscript{H}^k_{\dot\gamma a b} = 2 \dud 
{\bar\sigma} {ab} {\dot\alpha} {\dot\gamma} \bar D^k_{\dot\alpha} 
\mathscript{L},
\end{equation}
\begin{equation}
\mathscript{H}_{zab} =
- \frac{i}{2} \epsilon_{ij} \epsilon^{\alpha\gamma} 
\du \sigma {ab\gamma} \beta D^i_\alpha D^j_\beta \mathscript{L}
+ ik\epsilon_{ij} \epsilon^{\dot\beta\dot\gamma} 
{{{\bar\sigma}_{ab}\/}^{\dot\alpha}}_{\dot\gamma}
\tr(\bar \mathscript{W}_{\dot\alpha}^{i} \bar 
\mathscript{W}_{\dot\beta}^{j}),
\end{equation}
\begin{equation} 
\mathscript{H}_{\bar z ab} = \frac{i}{2} \epsilon_{ij} 
\epsilon^{\dot\alpha\dot\gamma} \dud {\bar\sigma} {ab} {\dot\beta} 
{\dot\gamma} \bar D_{\dot\alpha}^i \bar D_{\dot\beta}^j \mathscript{L} 
-ik\epsilon_{ij} \epsilon^{\beta\gamma} {{\sigma}_{ab\gamma}\/}^{\alpha}
\tr(\mathscript{W}_{\alpha}^{i} \mathscript{W}_{\beta}^{j}),
\end{equation}
\begin{equation}
\mathscript{H}_{a z\bar z} = - \frac{i}{4} \epsilon_{ij} \du {\bar\sigma} a 
{\dot\alpha\alpha} [D^i_\alpha, \bar D^j_{\dot\alpha}] \mathscript{L} + 2 i 
k \epsilon_{ij}  \du {\bar\sigma} a {\alpha\dot\alpha} \tr 
(\mathscript{W}^i_\alpha \bar \mathscript{W}^j_{\dot\alpha}) ,
\end{equation}
\begin{equation}
\mathscript{H}_{abc} = -\tfrac18 \epsilon_{abcd}\bar \sigma^{d \dot\beta 
\alpha}
\epsilon_{ij}( [D^i_\alpha, \bar D^j_{\dot\beta}] \mathscript{L}
+ 8 \kappa \tr(\mathscript{W}_{\alpha}^{i} \bar 
\mathscript{W}_{\dot\beta}^{j})).
\end{equation}
Furthermore, superfield $\mathscript{L}$ is constrained to satisfy
\begin{equation}
D^{(i}_\alpha \bar D^{j)}_{\dot\beta} \mathscript{L} = -2\kappa \tr(
\mathscript{W}_{\alpha}^{(i} \bar \mathscript{W}_{\dot\beta}^{j)})
\label{z1}
\end{equation}
as well as 
\begin{equation}
\epsilon^{\alpha\beta} D^{(i}_{\alpha} D^{j)}_{\beta} \mathscript{L} =
4\kappa 
\tr(\bar\mathscript{W}^i_{\dot\alpha}\bar\mathscript{W}^{j\dot\alpha}),
 \qquad \epsilon^{\dot\alpha\dot\beta} \bar D^{(i}_{\dot\alpha}
\bar D^{j)}_{\dot\beta} \mathscript{L} = 4\kappa \tr(\mathscript{W}^{i 
\alpha}
\mathscript{W}^{j}_{\alpha})
\label{z2}
\end{equation}
It follows that the addition of the Chern-Simons terms modifies both the 
solutions of the Bianchi identities and the constraints on $\mathscript{L}$.
In particular, the constraints now include terms proportional to 
$\mathscript{W}_{\alpha}^{(i} \bar \mathscript{W}_{\dot\beta}^{j)}$ and
$\mathscript{W}^{i \alpha}\mathscript{W}^{j}_{\alpha}$ and its conjugate. 
However,
these constraints can be rewritten to be more closely analogous with 
\eqref{z3} and \eqref{z4}. Using \eqref{x4} and the fact that
$\bar \mathscript{D}^{i}_{\dot\alpha}\mathscript{W}=
\mathscript{D}^{i}_{\alpha}\bar\mathscript{W}=0$, it is straightforward 
to show that \eqref{z3} and \eqref{z4} can be written in the form
\begin{equation}
D^{(i}_{\alpha} \bar D^{j)}_{\dot\beta} \Sigma=0
\label{z5}
\end{equation}
and
\begin{equation}
\epsilon^{\alpha \beta} D^{(i}_{\alpha} D^{j)}_{\beta} \Sigma+
\epsilon^{\dot\alpha \dot\beta} \bar D^{(i}_{\dot\alpha} \bar D^{j)}
_{\dot\beta} \Sigma=0
\label{z6}
\end{equation}
respectively, where
\begin{equation}
\Sigma=\mathscript{L}-2\kappa \tr(\mathscript{W}+\bar\mathscript{W})^{2}
\label{z7}
\end{equation}
As was discussed in \cite{PLB-392-85}, and used to construct the superfield 
action for the 
pure vector-tensor action \eqref{zz}, the generic form for a supersymmetric
action in central charge superspace is necessarily of the form
\begin{equation}
S=\int{d^{4}x\{D^{\alpha}_{i}D_{\alpha j}-\bar D_{\dot\alpha i} \bar
D^{\dot\alpha}_{j}\}M^{ij}}
\label{z8}
\end{equation}
where $M^{ij}$ must be real, symmetric in the indices $i$ and $j$ and
satisfy
\begin{equation}
D^{(i}_{\underline{\alpha}}M^{jk)}=0
\label{z9}
\end{equation}
For example, for the pure vector tensor multiplet discussed in Section 3,
we chose $M^{ij}=-W^{\beta (i}W^{j)}_{\beta}+\bar W_{\dot \beta}^{(i}
\bar W^{j)\dot\beta}$. In the case of the Chern-Simons modified theory,
the appropriate action is given by \eqref{z8} where
\begin{equation}
M^{ij}=D^{\beta (i}\Sigma D^{j)}_{\beta}\Sigma+\Sigma 
D^{\beta (i}D^{j)}_{\beta}\Sigma-\bar D^{(i}_{\dot\beta}\Sigma \bar
D^{j) \dot\beta}\Sigma
\label{z10}
\end{equation}
Note that this expression satisfies the above criteria. It is 
straightforward to expand this action into component fields. We will do
this, and explore other issues involving theories of this type, in future 
publications. 

\section*{Acknowledgments}

As this paper was being written, two related publications appeared. The 
first of these \cite{HEP-TH/9706108} discussed the two-form formulation of 
the vector-tensor multiplet and its coupling to Chern-Simons terms. The 
second paper \cite{HEP-TH/9706169} presented related topics within the 
context of harmonic superspace. We would like to thank Daniel Waldram for 
many important conversations.

This work was supported in part by DOE Grant No.\ DE-FG02-95ER40893, the 
University of Pennsylvania Research Foundation and the Alexander von 
Humboldt Foundation.

\providecommand{\bysame}{\leavevmode\hbox 
to3em{\hrulefill}\thinspace}

\end{document}